\documentclass[preprint]{aastex}
\usepackage{graphicx}
\usepackage{natbib}
\usepackage{amsmath}
\usepackage{bm}
\begin{document}

\title{A new pattern in Saturn's D ring created in late 2011}
\author{M. M. Hedman$^a$ and M.R. Showalter$^b$}
\affil{ $^a$Department of Physics, University of Idaho, Moscow ID 83844-0903\\
$^b$ SETI Institute, Mountain View CA 94043}

\maketitle

Images obtained by the Cassini spacecraft between 2012 and 2015 reveal a periodic brightness variation in a region of Saturn's D ring that previously appeared to be rather featureless. Furthermore, the intensity and radial wavenumber of this pattern have decreased steadily with time since it was first observed. Based on analogies with similar structures elsewhere in the D ring, we propose that this structure was created by some event that disturbed the orbital motions of the ring particles, giving them finite orbital eccentricities and initially aligned pericenters. Differential orbital precession then transformed this structure into a spiral pattern in the ring's optical depth that became increasingly tightly wound over time. The observed trends in the pattern's radial wavenumber are roughly consistent with this basic model, and also indicate that the ring-disturbing event occurred in early December 2011. Similar events in 1979 may have generated the periodic patterns seen in this same region by the Voyager spacecraft. The 2011 event could have been caused by  debris striking the rings, or by a disturbance in the  planet's electromagnetic environment. The rapid reduction in the intensity of the brightness variations over the course of just a few years indicates that some process is either damping orbital eccentricities in this region or causing the orbital pericenters of particles with the same semi-major axis to become misaligned. 

\section{Introduction}

Saturn's D ring lies between the planet and its main ring system, and while it is very faint, it is also among the most dynamic of Saturn's rings.  In particular, the outer D ring (73,200-74,500 km from Saturn's center) contains periodic brightness variations with time-variable wavelengths that were probably generated by an event in 1983 that abruptly caused all the ring particles to follow inclined and eccentric orbits with initially aligned ascending nodes and pericenters \citep{Hedman07, Hedman11, Hedman15}.  Similar structures in Jupiter's main rings appear to have formed in 1994, when the remnants of comet Shoemaker-Levy 9 were crashing into Jupiter, and so were likely generated by cometary debris impacting Jupiter's rings  \citep{Showalter11}. The patterns in Saturn's rings could have been created in a similar fashion, although there are no direct observations of cometary material striking Saturn's rings in the early 1980s.

\begin{figure}
\resizebox{6in}{!}{\includegraphics{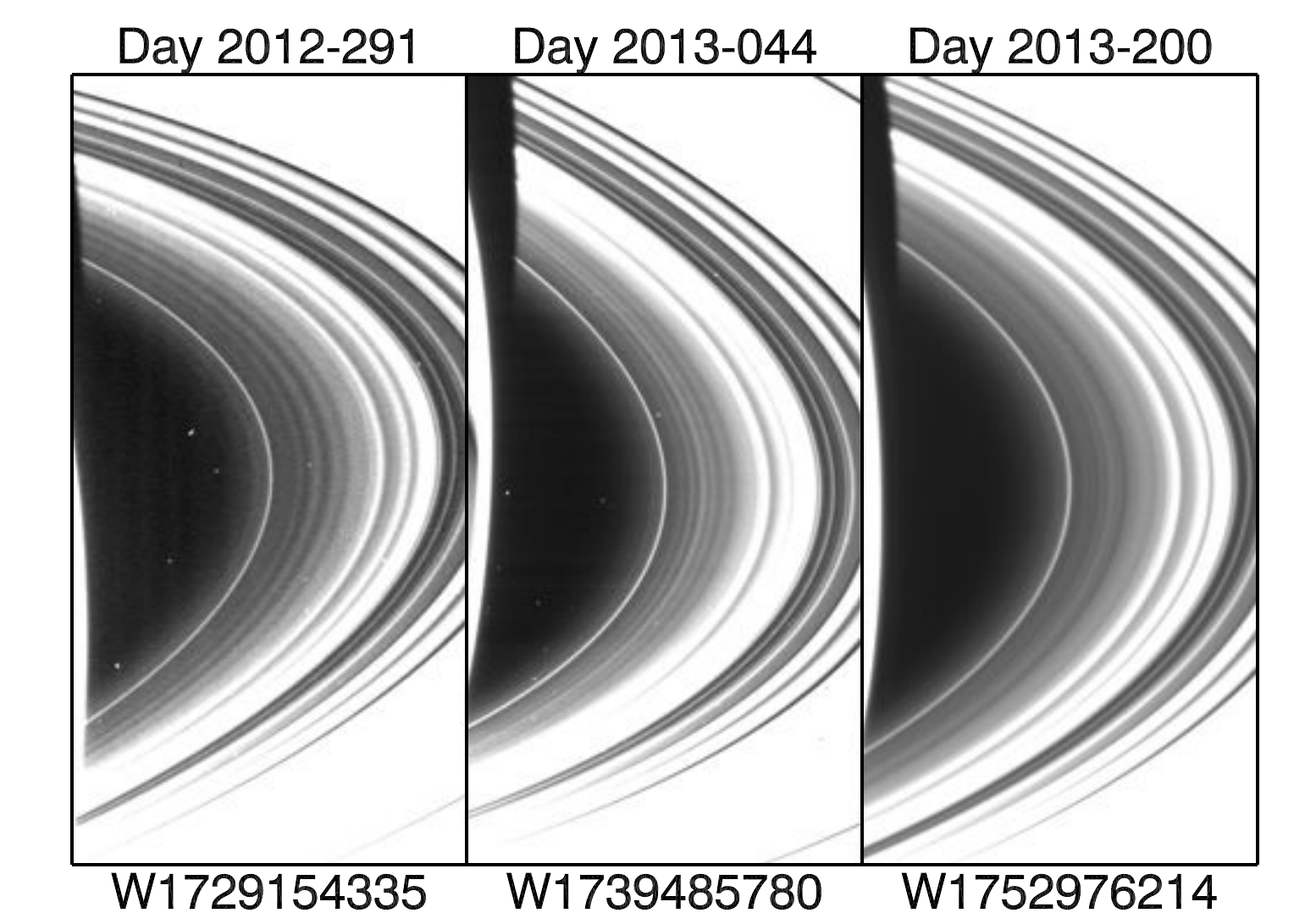}}
\caption{Three images of Saturn's D ring that show a periodic brightness variation in the relatively dark region outside the innermost narrow ringlet. All three images were obtained by Cassini's Wide Angle Camera at phase angles of 176.5$^\circ$, 175.6$^\circ$ and 175.4$^\circ$, respectively. The first and third images were obtained through the camera's RED filter, whereas the middle image was captured using the clear filters \citep{Porco04}. While the images have been cropped and rotated to facilitate comparisons, they all use a common stretch.}
\label{newpatim}
\end{figure}

Recent data from the Cassini spacecraft now reveal that a more recent event has disturbed some of the material in the inner D ring. Specifically, images obtained since 2012 of the region between 68,000 and 70,000 km from Saturn's center reveal a new periodic pattern with a time-variable wavelength (see Figure~\ref{newpatim}). The wavelength of this pattern has steadily decreased over time since this pattern first appeared, much like the structures that were created in Jupiter's rings in 1994 and Saturn's  rings in 1983. Indeed, the observed trends in this new pattern's wavelength are roughly consistent with differential apsidal precession in Saturn's gravitational field, and so match the expected evolution of a pattern created by yet another event that caused the ring particles to acquire finite orbital eccentricities with initially aligned pericenters.  However, unlike other examples of this type of structure, which were observed years or decades after they formed, this new pattern appears to have been created in late 2011, just seven months before it was first imaged. This pattern also differs from previously examined structures in that it seems to be short-lived, becoming nearly invisible in just a few years.

At present, it is still unclear what exactly created this new structure in the rings and what is causing the pattern to dissipate so rapidly. However, since Cassini was in orbit around Saturn when the ring was disturbed,  other instruments onboard Cassini might be able to shed light on what exactly happened in 2011. In order to facilitate these efforts, this report describes the currently-available data on this pattern and what information they provide about the pattern's origin and evolution. Section~\ref{theory} reviews the theory behind such time-variable spiral patterns, while Section~\ref{obs}  describes the relevant Cassini observations of the new structure and how they were processed to obtain estimates of the pattern's amplitude and wavelength. Section~\ref{results} then examines the trends in these parameters. Section~\ref{voyager} compares this new pattern to similar structures observed by the Voyager spacecraft in the early 1980s that might be due to earlier ring-disturbing events.  Finally, Section~\ref{discussion} discusses how the ring could have been disturbed and what might be responsible for the short damping times. 

\section{Theoretical background}
\label{theory}

Most of the previously studied periodic brightness variations with steadily decreasing wavelengths have been interpreted as vertical corrugations \citep{Hedman07, Hedman11, Showalter11}. Such corrugations appear as brightness variations because the warped surface places different amounts of material along different lines of sight. However, the pattern illustrated in Figure~\ref{newpatim} cannot be attributed to a corrugation. Brightness patterns generated by vertical corrugations are expected to fade out and undergo ``contrast reversals" near the ring ansa, where the camera's line of sight is nearly parallel to the local slopes. By contrast, the patterns in Figure~\ref{newpatim} have nearly the same intensity throughout the image, indicating that they are variations in the local surface density (see also Appendix A). \citet{Hedman15}  identified similar periodic  modulations with steadily decreasing wavelengths in the local surface density of the outer D ring that probably arose from an event that gave the ring particles finite orbital eccentricities and aligned orbital pericenters. The theory behind such ``eccentric spiral'' patterns is developed in \citet{Hedman15}, but we briefly summarize the relevant calculations here for reference.

Imagine that at some time $t_i$ all the particles in this part of the rings had orbits with finite eccentricity $e$ and the same pericenter longitude $\varpi$ (which can be set equal to zero). Particles orbiting at different semi-major axes $a$ will have different precession rates $\dot{\varpi}(a)$, so at any later time $t_f$ the pericenter location  will  also be position-dependent: $\varpi(a,t_f)=(t_f-t_i)\dot{\varpi}(a)$.  As long as Saturn's oblateness dominates the precession rates, $\dot{\varpi}(a)$ will be a monotonically decreasing  function of $a$, and the pericenter location will become an increasingly tightly wrapped trailing spiral.  Such trends in the pericenter location naturally give rise to variations in the ring's surface density. 

So long as the particles at any given semi-major axis follow the same trajectory (called a streamline), the local surface particle number density $\sigma$ is inversely proportional to the radial distance between adjacent streamlines:
\begin{equation}
\sigma=\frac{\sigma_o}{\partial r/ \partial a},
\label{streamline}
\end{equation}
where $\sigma_o$ is the average surface particle number density, and $r$ is the radial position of the streamline as a function of longitude $\theta$:
\begin{equation}
r=a-ae\cos(\theta-\varpi).
\end{equation}
In the vicinity of any given semi-major axis $a_o$, the pericenter location can be well approximated by the first two terms of a Taylor expansion:
\begin{equation}
\varpi(a,t_f)=\varpi(a_o,t_f)+\frac{\partial\dot{\varpi}}{\partial a}(t_f-t_i)(a-a_o).
\label{peria}
\end{equation}
In this situation the streamline equation becomes
\begin{equation}
r=a-A\cos(\theta-\varpi(a_o,t_f)+k(a-a_o)),
\end{equation}
where $A=ae$ is the magnitude of the radial excursions and 
\begin{equation}
k=\left|\frac{\partial\dot{\varpi}}{\partial a}\right|(t_f-t_i)
\label{eqkr}
\end{equation}
is their  wavenumber. Provided $A$ and $k$ are approximately constant on radial scales of order $k^{-1}$, then the radial derivative of $r$ is approximately:
\begin{equation}
\frac{\partial r}{\partial a}=1+Ak\sin[\theta-\varpi(a_o,t_f)+k(a-a_o)].
\end{equation}
As long as $Ak<<1$  this expression, combined with Equation~\ref{streamline}, implies that the surface number density of this perturbed ring is approximately:
\begin{equation}
\sigma\simeq\sigma_o\left[1-Ak\sin[\theta-\varpi(a_o,t_f)+k(a-a_o)]\right].
\end{equation}
For a vertically-thin, low-optical-depth structure like the D ring, the optical depth $\tau$ is proportional to the particle number density $\sigma$, so the fraction optical depth variations due to this pattern are:
\begin{equation}
\frac{\delta\tau}{\tau}\simeq-Ak\sin[\theta-\varpi(r_o,t_f)+k(r-r_o)].
\label{drho}
\end{equation}
(Note that the semi-major axes $a$ and $a_o$ have now been replaced  with the radii $r$ and $r_o$; this is valid so long as $Ak<<1$). The ring therefore exhibits a spiral pattern in its opacity, such that at a given longitude $\theta$, the ring's optical depth varies periodically with an amplitude $Ak$ and wavelength $2\pi/k$. 

Saturn's finite oblateness, quantified by the gravitational harmonic $J_2$,  is the dominant force driving apsidal precession in the D ring \citep{Hedman14, Hedman15}. Hence, to first order (and assuming $a \simeq r$), the precession rate should be given by the following expression
\citep{MurrayDermott}:
\begin{equation}
\dot{\varpi}=\frac{3}{2}nJ_2\left(\frac{R_s}{r}\right)^2
\end{equation}
where $R_s=60,330$ km is the fiducial planetary radius used in the calculation of $J_2$ and $n=\sqrt{GM/r^3}$ is the zeroth-order estimate of the local mean motion ($G$ being the gravitational constant and $M$ being Saturn's mass). This yields the following expression for the pattern's wavenumber:
\begin{equation}
k=\frac{21}{4}J_2\sqrt{\frac{GM}{r^5}}\left(\frac{R_s}{r}\right)^2(t_f-t_i).
\label{keq}
\end{equation}
Note that non-gravitational forces and higher-order terms in Saturn's gravity field can alter the relevant precession rates and  cause the actual wavenumber to deviate from the above prediction. However, the above expression is still a reasonable first-order approximation of the trends one  should expect to see for a spiral pattern that arose from an event that caused the ring particles to suddenly acquire finite orbital eccentricities. 

\section{Observations and Data Reduction}
\label{obs}

\subsection{Observations included in analysis}

The pattern visible in Figure~\ref{newpatim} has a wavelength that ranges between 100 and 1000 km, so it can be easily observed even in relatively low-resolution images. However, the part of the ring that contains this pattern is very strongly forward-scattering, and so the periodic brightness variations can only be detected clearly in images obtained at sufficiently high phase angles. Thus we performed an exhaustive search for images of this region taken between 2011 and early 2015 at phase angles higher than 130$^\circ$. This search included images obtained by both the Narrow Angle Camera (NAC) and the Wide Angle Camera (WAC) components of the Imaging Science Subsystem onboard Cassini \citep{Porco04, West10}. When multiple images were taken of the same region, the images taken through the relevant camera's clear filters were preferred because they  had the best signal-to-noise. However, two observation sequences obtained while the spacecraft flew through Saturn's shadow did not include any clear-filter images. Since these exceptionally high-phase observations contained some of the clearest images of these structures, the relevant images obtained through the WAC's RED filter were used in the main part of this analysis.\footnote{A few images using different filters are examined below in the context of the pattern's amplitude variations (see Figure~\ref{colcomp}), but we did not attempt a full analysis of these images because their lower resolution made the estimated wavelengths more uncertain.} The RED filter images were selected not only because they had the best resolution, but also because the effective wavelength of the RED filter is 647 nm, which is close to the effective wavelengths of the NAC and WAC clear filters (651 nm and 634 nm, respectively, Porco {\it et al.} 2004). The brightness measurements through these various filters should therefore be reasonably comparable to each other.

This search yielded ten observing sequences where the D-ring pattern was clearly detectable. Six of these sequences contained observations of a single ring ansa while the other four observed both ansae. Treating the data from each ansa separately (see below) gives a total of fourteen observations, whose properties are summarized in Table~\ref{obstab}. Note that while some of these observations consist of a single image, others consist of a sequence of multiple images. In these sequences, the images considered for this analysis had the best combination of resolution and signal-to-noise (these sets also excluded any image containing obvious imaging artifacts from cosmic rays that would corrupt the brightness profiles).

\subsection{Preliminary data reduction}

Each image was calibrated using the standard CISSCAL calibration routines \citep{Porco04, West10}, which remove instrumental backgrounds, flat-field the image and convert the raw data numbers into $I/F$, a measure of surface reflectance that is unity for a Lambertian surface viewed and illuminated at normal incidence.\footnote{See also {\tt http://pds-rings.seti.org/cassini/iss/calibration.html}} Each image was geometrically navigated using either stars in the field of view or the position of the C-ring's (assumed circular) inner edge.

\begin{table}[tbp]
\caption{Summary of available observations}
\label{obstab}
\resizebox{6.5in}{!}{\begin{tabular}{|c|c|c|c|c|c|c|}\hline
Images$^a$ & UTC & Phase & $\lambda$(69,000 km) & $A\varpi_0P(\alpha)\tau/4$ & Amplitude \\
 & & & (km) & (m) & (km) \\ \hline
N1719551308-N1719564388(25) & 2012-180T06:05:11 & 148.4$^\circ$ & 1076.9 & 0.1219 & --- \\
N1719549271(1) & 2012-180T03:42:15 & 154.9$^\circ$ & 1080.4 & 0.2552 & --- \\
W1721658117(1) & 2012-204T13:29:28 & 169.2$^\circ$ &  780.3 & 0.3314 &   6.29 \\
N1728999854-N1729023605(30) & 2012-289T16:08:50 & 143.4$^\circ$ &  581.7 & 0.0538 & --- \\
W1729153528(1)$^b$ & 2012-291T07:32:11 & 174.2$^\circ$ &  604.3 & 0.7730 &   5.15 \\
W1729154335(1)$^b$ & 2012-291T07:45:38 & 176.5$^\circ$ &  634.5 & 1.6930 &  11.72 \\
W1739485780(1) & 2013-044T21:35:18 & 175.6$^\circ$ &  429.1 & 1.4826 &   6.16 \\
W1752974769(1)$^b$ & 2013-201T00:30:21 & 177.9$^\circ$ &  332.1 & 3.0309 &   4.54 \\
W1752976214(1)$^b$ & 2013-201T00:54:24 & 175.4$^\circ$ &  330.6 & 0.3282 &   1.51 \\
N1761035999-N1761056255(25) & 2013-294T10:32:04 & 138.0$^\circ$ &  288.2 & 0.0056 & --- \\
N1765071135-N1765102855(62) & 2013-341T04:59:30 & 147.3$^\circ$ &  259.8 & 0.0096 & --- \\
N1770684650-N1770716969(20) & 2014-041T04:22:28 & 154.5$^\circ$ &  241.8 & 0.0085 & --- \\
N1770717591-N1770749910(20) & 2014-041T13:31:29 & 152.4$^\circ$ &  244.9 & 0.0133 & --- \\
N1802349729-N1802351228(2) & 2015-042T11:53:28 & 139.8$^\circ$ &  168.6 & 0.0013 & --- \\
\hline
\end{tabular}}
$^a$ Image names, followed by number of images used to derive amplitude and wavelength estimates in parentheses.

$^b$ Images obtained through the RED filter rather than the clear filters.
\end{table}

\begin{figure}
\centerline{\resizebox{5in}{!}{\includegraphics{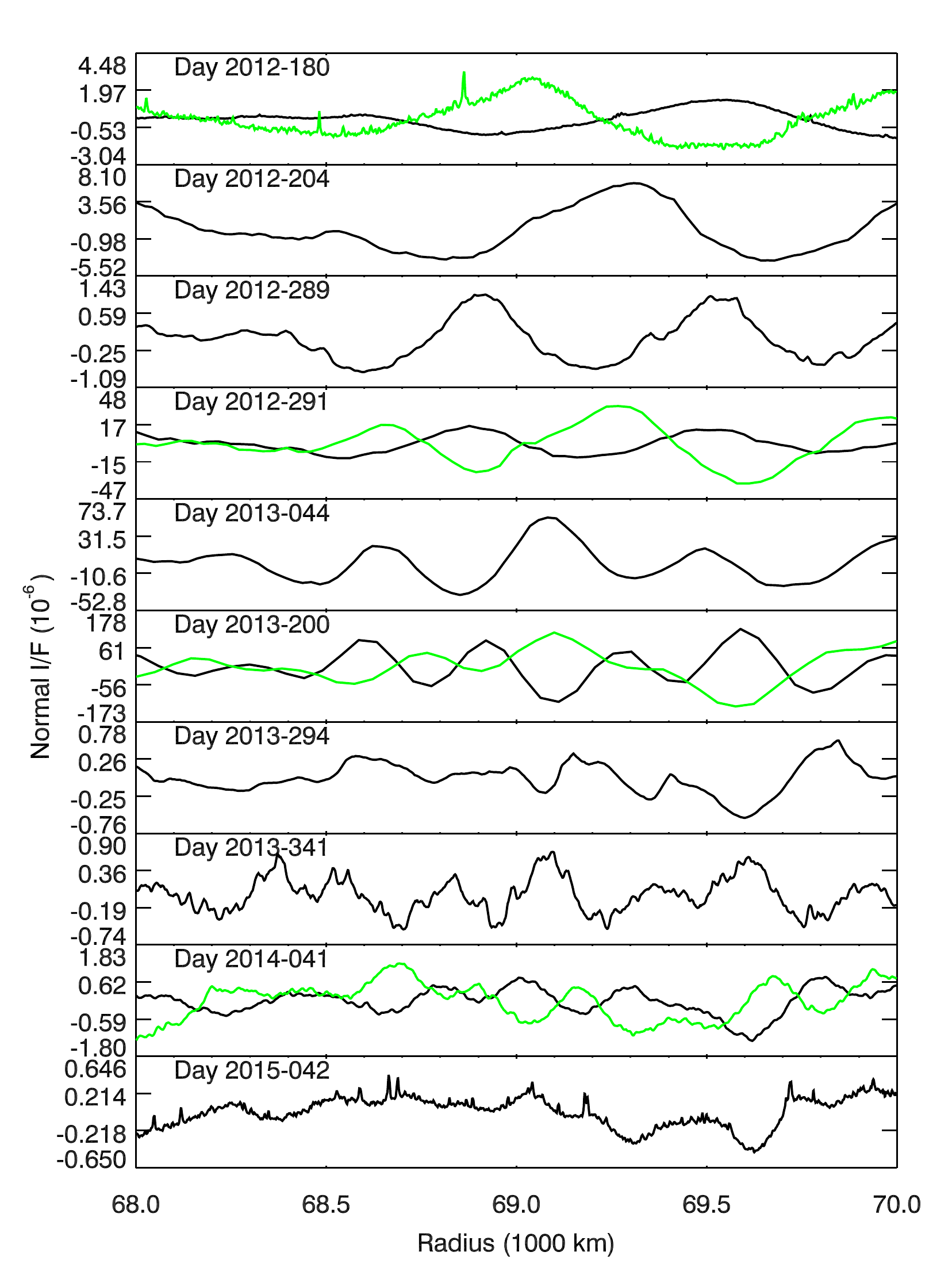}}}
\caption{The brightness profiles of the D ring considered in this analysis. Each panel shows one or two profiles of the ring's brightness versus radius. Each profile has had a quadratic trend removed to make the relevant brightness variations easier to see. When two profiles are shown in the same panel, these represent profiles from the two ring ansae. Note that whenever both ansae were observed, the maxima in one profile correspond to the minima in the other, and vice versa.}
\label{patcomp}
\end{figure}

Previous examinations of  periodic, time-variable patterns elsewhere in the D ring used brightness trends with longitude to isolate the signals due to an eccentric spiral from those due to vertical corrugations \citep{Hedman15}. However, these techniques did not reveal any clear signature of a vertical corrugation in the high-quality images of this new pattern (see Appendix A). Furthermore, over the limited range of longitudes visible in each image, the spiral nature of the pattern did not produce any obvious azimuthal trends. Hence, to simplify the analysis and improve signal-to-noise, the data from each image was reduced to a single profile of brightness versus radius (distance from Saturn's spin axis) by averaging over an appropriate range of longitudes (i.e., a region near the ring's ansa where the radial resolution was good and Saturn did not contaminate the ring signal). For a low optical depth ring like the D ring, inter-particle scattering and shadowing are negligible, and so the ring's apparent brightness is determined by the column density along the line of sight, which is inversely proportional to the sine of the ring opening angle. Hence we multiplied each profile by the sine of the ring opening angle to generate profiles of  ``normal $I/F$'', the brightness the ring would have if it was observed from precisely face-on.  Finally,  the profiles derived from images taken during a single observation of one ring ansa all showed the same brightness variations, so the relevant profiles were averaged together to further improve signal-to-noise. 

 These procedures yielded the fourteen brightness profiles illustrated in Figure~\ref{patcomp}.  In all of these profiles, periodic brightness variations are visible in the region between 68,000 and 70,000 km. In principle, this pattern could extend outside this region, but in practice it is difficult to isolate against the finer-scale structures found elsewhere in the D ring, Hence this investigation will focus exclusively on the region around 69,000 km where the relevant signal is easiest to detect and quantify. The wavelength of the brightness variations in this region clearly decreases over time, consistent with the trends shown in Figure~\ref{newpatim}, and with the expected behavior of an eccentric spiral. Furthermore, in all four cases  where both ring ansa were observed at the same time, the two profiles derived from opposite ansae are anti-correlated (maxima in one profile fall close to the minima in the other, and vice versa). This is consistent with Equation~\ref{drho}, which  predicts that the brightness variations on opposite sides of the ring (i.e., points where $\theta$ differs by $180^\circ$) will be $180^\circ$ out of phase with each other.


\subsection{Estimating pattern amplitudes and wavelengths}

Estimates of the patterns' amplitude and wavelength were derived using Fourier methods similar to those previously employed to quantify other periodic structures in Saturn's rings \citep{Hedman07, Hedman11, Showalter11}. More specifically, the procedures used here are similar to those \citet{Hedman15} employed to extract amplitudes and wavelength data from the (much shorter wavelength) eccentric spiral pattern in the outer D ring. The first step in this process is to take the radial derivative of the brightness profile, which suppresses large-scale background trends in the data and facilitates measurements of the periodic signal of interest here. Unlike other methods for high-pass filtering the data, taking the radial derivative has the advantage that  the amplitude of the pattern in the filtered profile is easily translated to the amplitude of the pattern in the original profile (If there is a periodic signal  with wavenumber $k$ and amplitude $\mathcal{A}$, then a periodic structure will exist in the radial derivative with the same wavenumber and an amplitude $\mathcal{A}k$.)

After computing the radial derivative of the brightness profile, the observed radii were translated into the rescaled distance parameter:
\begin{equation}
d=r\frac{2}{7}\left(\frac{r_o}{r}\right)^{9/2}, 
\end{equation}
where $r_o=$69,000 km. The ``rescaled wavenumber'' of a periodic structure measured in this transformed coordinate system is:
\begin{equation}
K=k\left(\frac{r}{r_0}\right)^{9/2},
\end{equation}
where $k$ is the true radial wavenumber. As discussed above, the wavenumber of an eccentric spiral pattern is expected to scale approximately as $r^{-9/2}$, and so the $K$ of such a pattern should be approximately constant.\footnote{Of course, higher-order corrections to the precession rates will introduce some weak trends in the rescaled wavelength. However, corrections due to higher-order terms in the planet's gravitational field are small enough ($<1\%$) to be ignored in this analysis.} The above transformation therefore enables a Fourier spectrum to be generated of the entire region between 68,000 km and 70,000 km, yielding precise wavenumber estimates. These estimates are also insensitive to variations in the amplitude of the pattern across the region, which can influence what part of the wave contributed most to the Fourier transform. Note that $K=k$ when $r=r_o=69,000$ km, so the wavenumbers obtained by this procedure can be regarded as estimates of the pattern's wavenumber at 69,000 km, near the middle of the region where the pattern is evident.

For each profile, an over-resolved Fourier spectrum of the rescaled radial-derivative data is computed by evaluating the Fourier transform for a tightly spaced array of $K$ values ($\delta K/K < 0.001$). These spectra contain a strong peak at the wavenumber of the desired periodic signal. This peak is fit to a Gaussian in order to obtain estimates of the pattern's wavenumber and amplitude.\footnote{Note that, in contrast to previous analyses by \citet{Hedman11, Hedman15}, for this investigation we fit a gaussian to the signal versus {\em wavenumber} rather than versus {\em wavelength} because the peaks in the Fourier spectra with the longest-wavelength patterns were noticeably asymmetric when plotted as functions of wavelength. In practice, the estimated wavenumbers derived from fitting the Fourier spectra as functions of wavenumber and wavelength only differed by at most 1\%.} Only data where the amplitude of the Fourier transform is at least half the peak value and within 10\% of the peak wavenumber are included in the fit. This fitted position of the peak in the Fourier spectrum then provides an estimate of  $K$, while the peak amplitude of the Fourier spectrum provides an estimate of the amplitude of the brightness variations $\mathcal{A}$ (after accounting for the factor of $k$ introduced by the taking the radial derivative).

While the wavenumber of the observed brightness variations simply equals the radial wavenumber of the eccentric spiral pattern, some more work is needed to translate the amplitudes of the brightness variations into estimates of the particles' average eccentricity. For sufficiently low optical depth rings, the normal $I/F$ ($\mu I/F$) is given by the expression:
\begin{equation}
\mu I/F =\frac{1}{4}\varpi_0P(\alpha)\tau,
\end{equation}
where $\varpi_0$ is the average geometric albedo of the ring particles, $P(\alpha)$ is the average phase function of the ring material, and $\tau$ is the ring's normal optical depth. The amplitude of the brightness variations is therefore directly proportional to the amplitude of the optical depth variations:
\begin{equation}
\mathcal{A} =\frac{1}{4}\varpi_0P(\alpha)\tau kA.
\end{equation}
Translating this information into an estimate of $A$ requires an estimate of the ring's mean brightness. Unfortunately, for most of the observations, it is difficult to estimate the average brightness between 68,000 and 70,000 km due to non-zero instrumental background signals. However, for the five highest-phase observations, the ring's signal-to-noise is higher and the relevant images also capture the region closer to Saturn, where the ring density is much lower (see Figure~\ref{newpatim}). Therefore, the mean brightness of the relevant ring region $\mathcal{S}$ can be estimated by subtracting the mean signal level between 62,000 and 64,000 km from the mean signal level between 68,000 and 70,000 km. In these cases, the amplitude of the radial motions in the eccentric spiral can be derived from the ratio $\mathcal{A}/\mathcal{S}k$. Table~\ref{obstab} provides these estimates of $A$, along with estimates of the pattern's radial wavelength (derived from the estimated $K$) and the parameter combination  $\mathcal{A}/k=\varpi_0P(\alpha)\tau A/4$ for all of the available observations. No formal error bars are provided for these values because systematic errors in the assumed background signal level will likely dominate the uncertainties in the amplitude estimates, and the wavelength estimates derived for these long-wavelength patterns can shift slightly due to interference from any other instrumental background trends and/or structures in this region. Since these sources of error are difficult to quantify {\it a priori}, the uncertainties in these parameters will be estimated from the scatter in the measurements about various trends.

\section{Results and trends in pattern parameters}
\label{results}

Table~\ref{obstab} lists the estimated amplitudes and wavelengths of the pattern derived from the fourteen available observations, while Figure~\ref{patpar} plots both the estimated wavenumber and amplitude of the pattern as functions of time. The wavenumber trends confirm that this pattern is likely an eccentric spiral produced by some ring-disturbing event like that described in Section~\ref{theory} above. The variations in the pattern's amplitude, however, were unexpected and imply that some other process is affecting the particles' orbital evolution.

\subsection{Trends in the pattern's wavenumber}

The wavenumber measurements shown in Figure~\ref{patpar} follow a very clear linear trend. Fitting this trend to a line, and using the scatter around the trend as an estimate of the $rms$ error in the estimates, yields a slope of  3.234$\pm$0.048$\times10^{-5}$ km$^{-1}$/day. For comparison, the expected winding rate for an eccentric spiral pattern, assuming currently-accepted values for Saturn's mass and gravitational harmonics \citep{Jacobson06} is 3.361$\times10^{-5}$ km$^{-1}$/day.\footnote{This number includes contributions from $J_2$ and higher-order gravitational harmonics. The first-order expression for the winding rate given in Equation~\ref{keq} says that the winding rate of the eccentric spiral at 69,000 km should be  $\dot{k}=2.782\times10^{-5}$ km$^{-1}$/day, assuming the currently-accepted values for Saturn's mass and $J_2$ \citep{Jacobson06}.}  This predicted rate is slightly faster (by between 2 and 3 standard errors) than the observed rate. However, the difference between the observed and expected rate is only few percent, so the evolution of this pattern's wavelength is very close to the expected behavior of an eccentric spiral. This quantitative match between observed and predicted winding rates, together with the anti-correlation between the profiles on opposite ring ansae mentioned in Section~\ref{obs} above, provides strong evidence that this new pattern is indeed an eccentric spiral.

\subsection{Trends in the pattern's amplitude}

\begin{figure}
\centerline{\resizebox{5in}{!}{\includegraphics{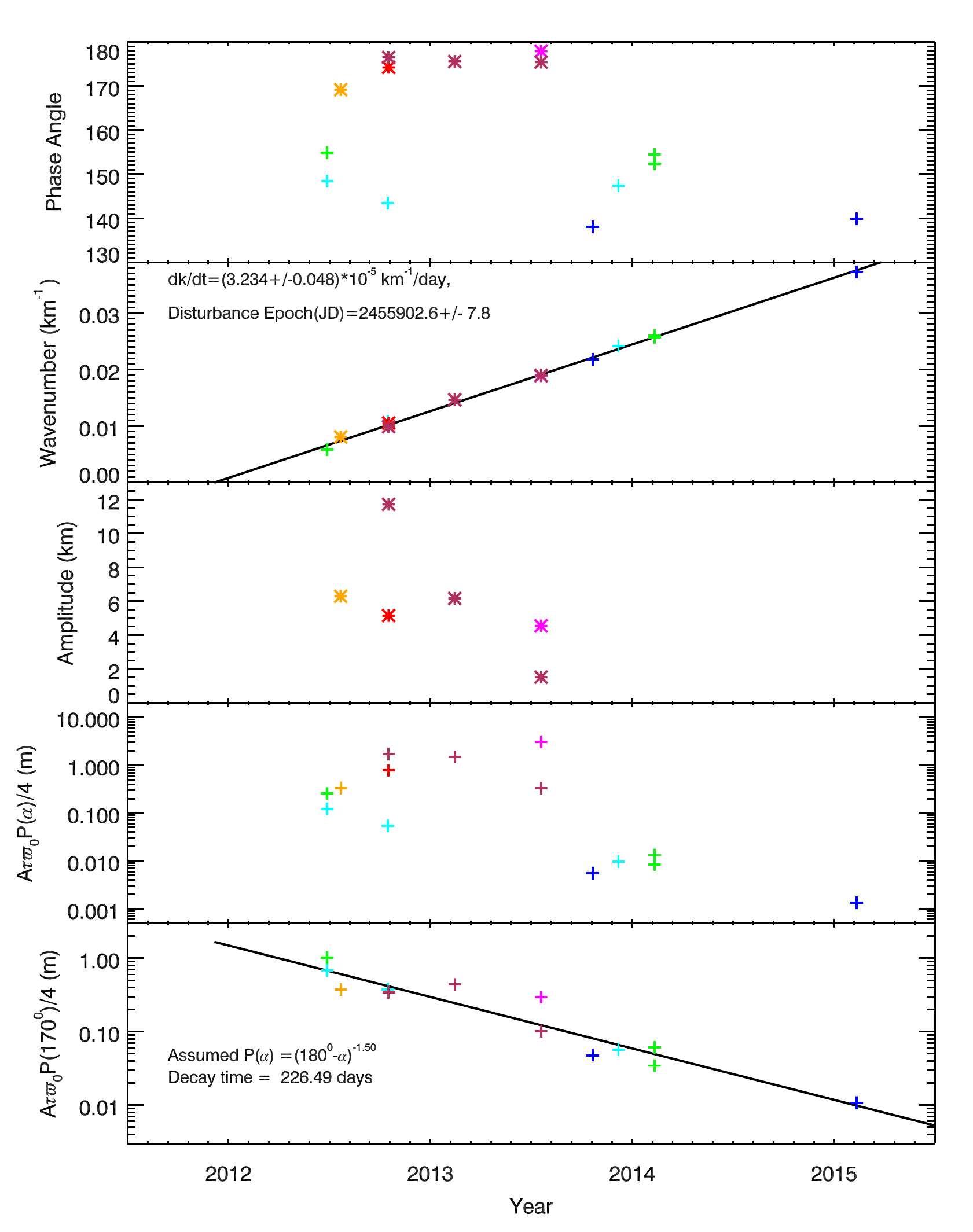}}}
\caption{Plots showing trends in the pattern's wavenumber and amplitude as functions of time. The top panel shows the phase angles of the relevant observations, with color codes used in the remaining panels. The second panel shows the pattern's wavenumber as a function of time. A linear trend in these data is evident. The third panel shows the estimated amplitudes of the radial motion, while the fourth panel shows the same amplitude $A$ times photometric factors (see text). In both these panels the pattern's amplitude clearly declines over time. The last panel scales the data in the previous panel by a power-law phase function to better illustrate the overall trend.}
\label{patpar}
\end{figure}

\begin{figure}
\resizebox{6in}{!}{\includegraphics{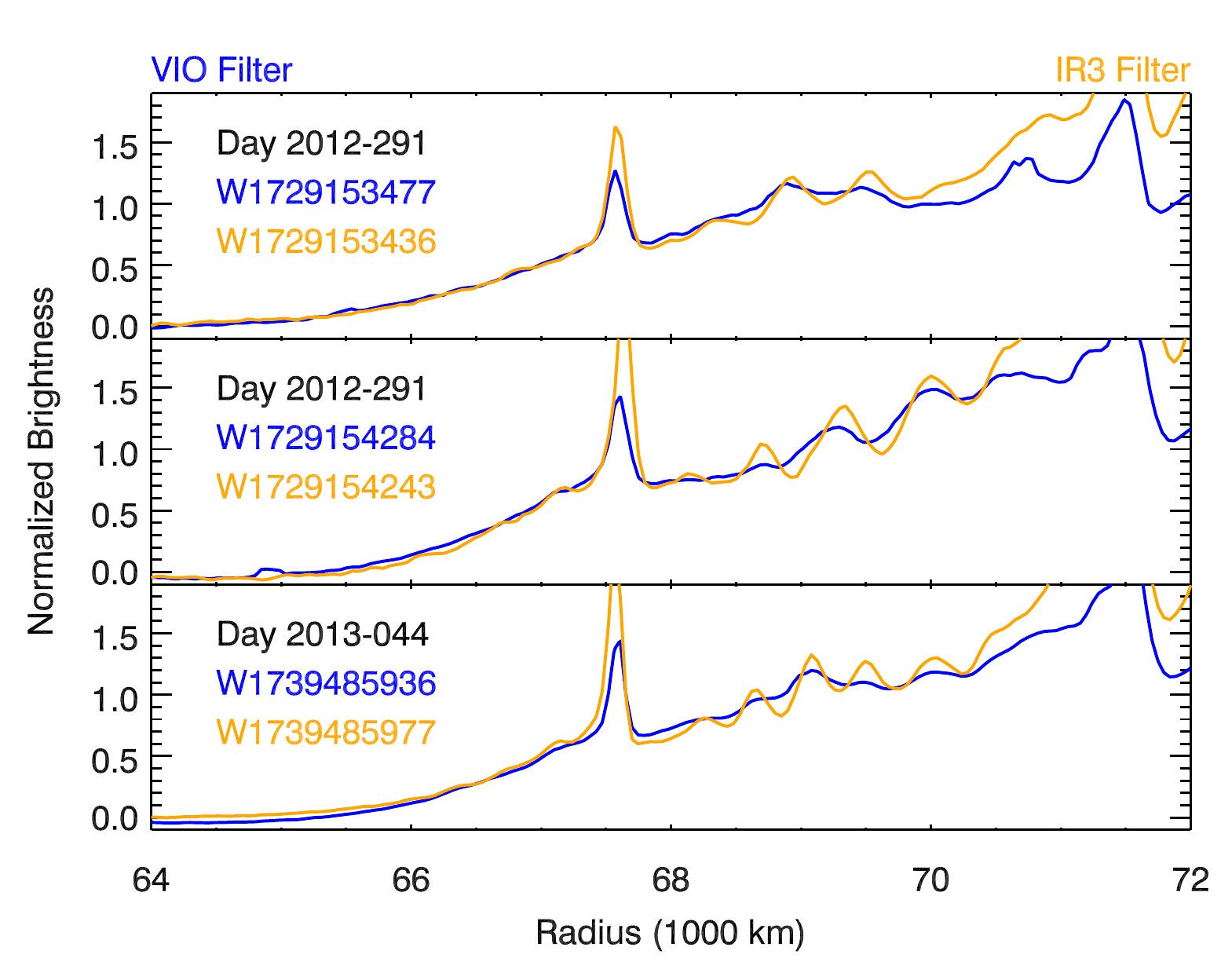}}
\caption{Comparisons of the pattern observed through the VIO (0.420 $\mu$m) and IR3 (0.916 $\mu$m) filters. Each panel shows profiles derived from a given observation sequence of one ring ansa at very high phase angles (174.5$^\circ$, 176.5$^\circ$ and  175.6$^\circ$, respectively). In all these profiles, the ring's brightness is normalized and offset so that it averages to zero between 62,000 and 64,000 km and averages to unity between 68,000 and 70,000 km.  In all three cases, the fractional amplitude of the pattern between 68,000 km and 70,000 km is clearly larger when viewed through the longer-wavelength IR3 filter than it is when viewed through the shorter-wavelength VIO filter.}
\label{colcomp}
\end{figure}

Surprisingly, the pattern's amplitude also shows variations both with time and with phase angle. For the six observations where $A$ can be estimated, the radial motions of the particles range between 1 and 12 km. These include two pairs of measurements made at nearly the same time but at different phase angles. In both cases, the observation made at the higher phase angle yields the bigger amplitude, so the particles that scatter light more efficiently at higher phase angles appear to have larger radial epicyclic motions. Since we are observing the rings at extremely high phase angles, the observed light is primarily scattered by diffraction, and so the scattering efficiency depends mostly on particle size. Larger particles scatter light over a smaller range of phase angles, so the increase in the pattern amplitude as the phase angle approaches 180$^\circ$ suggests that larger particles have larger organized eccentricities.

We can verify this supposition by comparing observations made at the same time through different broad-band filters, which are shown in Figure~\ref{colcomp}. Here we show profiles derived from images taken at the same time through two filters, the VIO filter with an effective central wavelength of 0.42 $\mu$m and the IR3 filter with an effective central wavelength of 0.92 $\mu$m \citep{Porco04}.  For each profile we subtract a constant background level so that the mean signal between 62,000 and 64,000 km is zero, and we normalize each profile so the average signal level between 68,000 km and 70,000 km is unity. In all of three cases, the pattern's fractional amplitude is clearly larger when viewed through the longer-wavlength IR3 filter than it is  when viewed through the shorter-wavelength VIO filter. This implies that the larger particles (which are more efficient at scattering at longer wavelengths) have larger average eccentricities than smaller particles, consistent with the observed trends with phase angle.

These synchronous amplitude variations must be taken into account when  comparing observations made at different times. Fortunately, there are two amplitude estimates derived from images taken at around 175.5$^\circ$ on Days 044 and 201 of 2013 (see Table~\ref{obstab}). During the 157 days between these observations, the amplitude of the particle's radial motions seems to have decreased by roughly a factor of 4. This same trend can also be observed in the various estimates of $A\tau \varpi_0P(\alpha)/4$ derived from the remaining observations. At any given phase angle, the photometric factors in that product should be constant, so any variations with time can be ascribed to changes in $A$. For example, the estimates of $A\tau \varpi_0P(\alpha)/4$ derived from the two observations obtained at phase angles of around 155$^\circ$ indicate that the amplitudes of the radial motions decayed by a factor of thirty over the course of the 592 days between Day 180 of 2012 and Day 41 of 2014.
 
A clearer picture of the temporal variations in the pattern's amplitude can be obtained by translating the estimates of $A\tau\varpi_0P(\alpha)/4$ into estimates of the scaled amplitude at a particular phase angle. In principle, this could be accomplished by computing a theoretical phase function for the debris in the ring using light-scattering codes based on Mie theory or some other photometric model. However, such a detailed  calculation is beyond the scope of this work because the effective phase function for a ring depends upon the particle size distribution, and an exhaustive analysis of the available spectral and photometric data would be needed to ascertain what range of particle size distributions could be present in the ring. Fortunately, at the high phase angles considered here the scattered light is primarily due to diffraction around individual ring particles, and in this limit the phase function of dusty systems can often be approximated with a simple power law in the scattering angle $180^\circ-\alpha$ \citep[e.g.][]{Hedman07, Hedman13}. In this particular case, comparing observations made at the same time but different phase angles indicates that the simple power-law phase function  $P(\alpha) =(180^\circ-\alpha)^{-1.5}$ provides a reasonable first-order approximation for the eccentric spiral pattern. This is somewhat steeper than the $P(\alpha) =(180^\circ-\alpha)^{-1.2}$ phase function that best fit the average brightness of the material \citep{Hedman07}, which is consistent with the above-mentioned trends in the pattern's amplitude with phase angle.

The bottom panel of Figure~\ref{patpar} presents the estimates of $A\tau\varpi_0P(170^\circ)/4$ derived from all the relevant images assuming the above phase function. While there is still substantial scatter in the data (probably because the assumed phase function is only a first-order approximation), it does appear that the amplitude of the radial motions in the eccentric motion is decaying exponentially with time. Fitting these data to a simple exponential model yields a decay time of around 220 days. This is about a factor of two slower than the decay rate that would be predicted based on the above comparisons of observations at 155$^\circ$ or 175.5$^\circ$ phase, but still indicates that the pattern's amplitude is rapidly decaying.

\section{Was a similar structure seen by Voyager?}
\label{voyager}

\begin{figure}[tb]
\resizebox{6.5in}{!}{\includegraphics{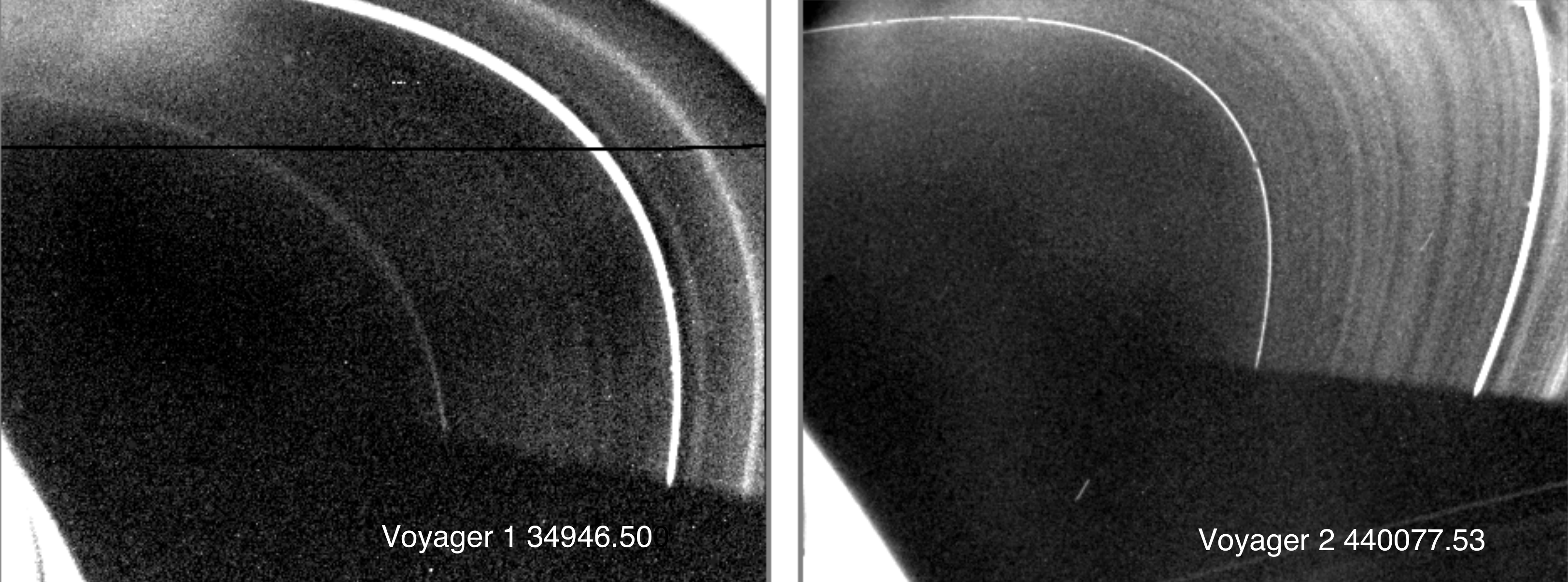}}
\caption{Voyager images of the D ring \citep [modified from][]{Showalter96}. The left-hand image was obtained by Voyager 1 in late 1980, and shows a faint periodic signature in the region between the narrow D68 ringlet and D72 (a bright, narrow ringlet that was the most prominent D-ring feature at this time). The right hand image was obtained by Voyager 2 in 1981, and shows a clearer pattern of brightness variations in this same region. }
\label{vgrfig}
\end{figure}

\begin{table}
\caption{Summary of Voyager observations}
\label{voytab}
{\begin{tabular}{|c|c|c|c|c|c|}\hline
Images & UTC & Phase & $\lambda$(69,000 km)  & Amplitude \\
 & & & (km)$^a$ &  (km)$^a$ \\ \hline
34946.50 & 1980-318T02:01:22 & 156.0$^\circ$ & 369-559 & 19-30  \\
44007.53 & 1981-238T05:12:47 & 164.0$^\circ$ & 242-276 & 24-24  \\
\hline
\end{tabular}}

$^a$ first estimate considers data from 68,000-70,000 km, while the latter uses 68,000-71,000 km.
\end{table}

This new pattern is reminiscent of the periodic brightness variations observed in this region by the Voyager spacecraft.  \citet{Showalter96} showed that each of the two Voyager spacecraft obtained one high signal-to-noise image of the D ring (see Figure~\ref{vgrfig}). In the Voyager 2 image, there is a clear periodic brightness variations extending between 68,000 and 71,000 km with a wavelength of around 300 km. In the earlier Voyager 1 image, the pattern is less clear but  appears to have a longer wavelength. 

Applying the above algorithms to these data yield estimates of the amplitude and the dominant wavelengths of these patterns. However, these patterns do not appear to be as periodic as the ones discussed above, so the wavelength estimates vary substantially depending on the radial range considered. Table~\ref{voytab} therefore provides two different wavelength estimates derived by considering the radial ranges of 68,000-70,000 km and 68,000-71,000 km (Note in both cases the reported wavelength is the value at 69,000 km). The estimated wavelength during the Voyager 1 epoch varies by 50\%, so precise analyses of the observed trends is not possible. However, it does appear that the wavelength of this pattern is declining with time at a rate between 3.12 and 4.03$\times10^{-5}$ km$^{-1}$/day, which would be consistent with an eccentric spiral pattern. 

Assuming that the pattern observed by Voyager is indeed an earlier eccentric spiral with the same winding rate as the new pattern seen by Cassini, the rings would need to have been disturbed 690-820 days before the (better measured) Voyager 2 observation, or between days 150 and 280 of 1979. One way to explain the irregularities of the pattern is that multiple patterns with slightly different wavelengths are overprinted on this region. Similar overprinted patterns have been identified in Jupiter's rings that could represent multiple passages of the comet Shoemaker-Levy 9 near the rings \citep{Showalter11}. 

The pattern observed by Voyager at Saturn also seems to have a substantially larger amplitude than the structure observed by Cassini in 2012-2013. For both Voyager images, the pattern would appear to have an amplitude of $\sim$25 km, much larger than any of the Cassini measurements (which were obtained at higher phase angles). 

It is also not obvious that the pattern's amplitude decayed between the Voyager 1 and 2 observations. The two observations yield comparable values of $A$,  but the later Voyager 2 image was obtained at a higher phase angle, which in the Cassini data yields a larger amplitude. Thus it could be that the amplitude of the pattern did decay, but that the phase angle variations between the two observations masked that decay. The pattern most likely did dissipate between the Voyager and Cassini epochs. Extrapolating the observed trends forward in time, we would expect a pattern with a wavelength of order 20 km to exist in this region during the Cassini mission,  and there is no evidence for such a structure in the available Cassini images. Thus the event that created the pattern seen in Cassini images from 2012-2015 is probably not unique, and similar events may have occurred in the last few decades.

\section{Discussion}
\label{discussion}

The above analysis indicates that (1) the new pattern arose from some event that pushed the D ring off-center in late 2011, (2) the resulting organized epicyclic motions have been steadily decaying over time, and (3) similar events may have occurred in 1979. In principle, the origin and evolution of the 2011 pattern (and possibly the 1979 pattern) could be quite different from those of the patterns created in 1983 because the 1983 event appears to have disturbed particles with a broad range of sizes in both the C and D rings \citep{Hedman07, Hedman11}, while the observable pattern generated by the 2011 event is predominantly made up of very small particles. Indeed, it is still not clear what is causing the pattern to dissipate or what happened to the rings in 2011 to create this structure in the first place. In light of these issues, this section discusses some processes that might be involved in both the creation and destruction of the new pattern, and therefore could be useful for future efforts to understand this structure.

\subsection{What caused the pattern to dissipate so rapidly?}

The trends in the pattern's amplitude indicate that the ring-particles' coherent epicylic motions are being damped on timescales shorter than a year. This could either be because the eccentricities of the particles' orbits are being damped, or because the pericenters of the particles at the same radial distance from the planet are becoming misaligned. We will consider both these options below.

Eccentricity damping is an attractive idea in this context because it appears to occur in other, nearby D-ring structures. In particular, the middle D ring (71,500-73,000 km from Saturn's center) contains structures that appear to be generated by Lindblad resonances with asymmetries in the planet's magnetosphere \citep{Hedman09res}. The morphology of these patterns depends on the eccentricity damping rate, and the observed structures are most consistent with a damping time of order a few weeks. However, such rapid orbital evolution does not appear to occur in other parts of the D ring. For example, the corrugations and eccentric spirals in the outer D ring (exterior to 73,200 km) are still visible decades after the ring-disturbing event \citep{Hedman07, Hedman15}. Also, a narrow ringlet located near the inner edge of the D ring around 67,630 km  has maintained a substantial eccentricity for  the entire Cassini mission \citep{Hedman14}. Thus the processes responsible for damping the ring-particles' orbital eccentricities would need to be most active between 68,000 km and 73,000 km. 

In principle, mutual collisions among the ring particles  at different semi-major axes could damp the particles' epicyclic motions. However, the opacity of this ring is so low that collisions among ring particles should be very infrequent. For low-optical depth rings the collision frequency for a given particle should be of order $3n\tau$ \citep{Schmidt09}, and in this part of the D ring this number corresponds to each particle experiencing one collision every couple of years, well below the observed dissipation rate. In principle, inter-particle collisions could be much more frequent if the
particles have net electric charges and interaction cross-sections much larger than their physical cross sections. However, even if the collision rate was enhanced by many orders of magnitude, the inter-particle interactions should only be able to attenuate the pattern if the particles with different pericenter locations can collide with each other, which will only happen when the amplitude of the particles' epicyclic motions is comparable to the pattern's radial wavelength. By contrast, the observed pattern appears to dissipate even when the pattern's amplitude is an order of magnitude smaller than its wavelength. 


Alternatively, the average orbital eccentricity of these small particles could be attenuated by drag forces from the ambient plasma and/or atmosphere. When a ring particle passes through a plasma or a neutral atmosphere, it will exchange momenta with the relevant ions and atoms, leading to a net force that opposes the ring particle's motion relative to the surrounding medium. If the particle is on an eccentric orbit, then this force will damp the particles' epicyclic motions on a timescale given by \citep{BHS01}:
\begin{equation}
T_d=\frac{s\rho_p \zeta}{\rho_a v}
\end{equation}
where $s$ is the particle radius, $\rho_p$ is the particle's internal density, $\rho_a$ is the average mass density of the plasma or neutral atmosphere, $v$ is the velocity of the particle relative to the ions and molecules, and $\zeta$ is a factor that is of order unity for cold plasma or a neutral atmosphere, but can be $\sim10^{-2}$ if the particle moves slowly relative to the mean thermal speed of the plasma ions. 

This mechanism for damping the amplitude of the new pattern is appealing because the damping timescale depends on the size of the particle, and this could naturally explain why the amplitude of the pattern appears to be larger at higher phase angles and longer optical wavelengths. Under these conditions, larger particles  should make a bigger contribution to the observed brightness, and since these particles will maintain their eccentricities for a longer period of time, the intensity of the pattern will be correspondingly larger. Furthermore, the particles in the region containing the new pattern appear to be smaller on average than those found in the outer D ring \citep{Hedman07}, which would explain why the new pattern is more transient.

Unfortunately, it is unlikely that the plasma or gas occupying the relevant parts of the D ring is dense enough to produce the observed attenuation of the pattern. Assuming the plasma/atmosphere co-rotates with the planet, then $v \simeq 9$ km/s. Furthermore, if the typical ring particles are ice-rich ($\rho_p \simeq 1$ g/cm$^3$), and of order 1 micron in radius \citep{Hedman07}, then a damping time or order 200 days for micron-sized particles would require  $\rho_a \sim 10^{-17} {\rm g/cm}^3 (\zeta s/1 \mu$m), which corresponds to a hydrogen number density of order $10^{6}/$cm$^3$ ($\zeta s/1 \mu$m). By comparison, the typical electron densities at the relevant altitudes above Saturn's cloudtops are  only around 10$^3$/cm$^3$ \citep{Kliore09, Nagy09}. Thus, unless the neutral density greatly exceeds the ion density, or the molecular density within the rings is very different from that elsewhere around the planet, plasma or atmospheric drag would appear to be far too weak to produce the required orbital evolution. 
 
We have been unable to identify any other physical process that could damp orbital eccentricities sufficiently fast to explain both the dissipation of the new pattern and the morphology of the resonantly-driven features elsewhere in the D ring. Hence we must consider the alternative possibility that  the pattern's apparent amplitude is decreasing with time because the orbital pericenters of particles with the same semi-major axes are becoming misaligned. The most straightforward way this could happen is if the apsidal precession rate of the ring varies with particle size because each particle has a finite, size-dependent electric charge that enables the planet's magnetic field to perturb their orbital motions (we thank a reviewer for suggesting this idea).  \citet{Hamilton93} derived analytical expressions for the orbital evolution of charged particles in orbit around Saturn, and if we combine Equations (30) and (31) from that paper (neglecting terms  due to solar radiation pressure), we obtain the following expression for the apsidal precession rate:
\begin{equation}
\dot{\varpi}=n\left[\frac{3J_2R_s^2}{2a^2(1-e^2)^2}+\frac{2nL}{\Omega_P(1-e^2)^{3/2}}\right],
\label{precl}
\end{equation}
where $R_s=60,330$ km, $J_2=0.017$ and $\Omega_P\sim1.7\times10^{-4}/$s are Saturn's radius, quadrupole gravitational moment, and spin rate, while $a, n$ and $e$ are the particle's (unperturbed) mean motion, semi-major axis, and eccentricity. For the purposes of this study, we may assume that $a \simeq 69,000$ km, $n\sim3.4\times10^{-4}/$s and that the particles' eccentricity are sufficiently small that $(1-e^2)\simeq 1$. The parameter $L$ is a measure of the particles' sensitivity to electromagnetic perturbations and is given by the following expression \citep{Hamilton93, Hamilton93e}:
\begin{equation}
L=\frac{q}{m}\frac{g_{1,0}R_s^3\Omega_P}{GM}
\end{equation}
where $g_{1,0} \sim 21.2 \mu$T is the dipole Gauss coefficient  of Saturn's magnetic field,  $M=5.7\times 10^{26}$ kg is Saturn's mass, and $q/m$ is the particle's charge-to-mass ratio.  Assuming the ring particles are compact and ice-rich grains, $q/m$ can be expressed in terms of the particle's electrostatic potential $\Phi$ and radius $s$ \citep{Hedman12}, yielding $L=-9.6\times10^{-4}(\Phi/-2 $V$)(s/1 \mu$m$)^{-2}$.  Inserting all the relevant numerical factors yields the (size-dependent) precession rate:
\begin{equation}
\dot{\varpi}=33^\circ/{\rm day}\left[1-0.2\frac{(\Phi/-2{\rm V})}{(s/1 \mu{\rm m})^2}\right].
\end{equation}
The relevant ring particles could have a range of different densities, shapes, and charge retention properties that would affect their charge-to-mass ratios. However, for the sake of simplicity we may consider a scenario where the ring particles all are composed primarily of compact water-ice \citep{Hedman07} and have similar potentials of order -2 V \citep[i.e. similar to the E ring,][]{Kempf06}. In this limit, two particles with the same semi-major axis but different radii $s_1$ and $s_2$ will have precession rates that differ by:
\begin{equation}
\delta \dot{\varpi}=(7^\circ/{\rm day})\frac{\Phi}{-2{\rm V}}\left|\frac{(s_1^2-s_2^2)(1 {\rm \mu m})^2}{s_1^2s_2^2}\right|.
\end{equation}
Thus the timescale required for the particles' orbital pericenters to become misaligned by more than 1 radian is
\begin{equation}
t_{\delta \varpi}=\frac{1}{\delta \dot{\varpi}}=({\rm 8  days})\frac{-2{\rm V}}{\Phi}\left|\frac{s_1^2s_2^2}{(s_1^2-s_2^2)(1 {\rm \mu m})^2}\right|.
\end{equation}
For a ring consisting of many particles with a distribution of sizes, the organized eccentric spiral pattern will be attenuated over a timespan equivalent to the appropriately weighted average of the above timescale, which can be written as:
\begin{equation}
\langle t_{\delta \varpi}\rangle =({\rm 8  days})\frac{-2{\rm V}}{\Phi}\left(\frac{s_{\rm diff}}{1 {\rm \mu m}}\right)^2.
\end{equation}
where $s_{\rm diff}$ is the average particle size of the visible particles times a coefficient of order unity that depends on the exact shape of the size distribution.  The observed dissipation timescales of order 200 days implies that $s_{\rm diff} \simeq {\rm 5 \mu m} \sqrt{\Phi/(-2 {\rm V})}$, which is consistent with the typical ring particles being charged to a potential of a few volts and of order a few microns wide.

This estimate of $s_{\rm diff}$ may also be compatible with the measured value of the pattern's winding rate. Recall that the winding rate is equivalent to the radial gradient of the precession rate, and so the size-dependent term in Equation~\ref{precl} will produce a similar correction to the winding rate:
\begin{equation}
\left|\frac{\partial \dot{\varpi}}{\partial a}\right|=\frac{n}{a}\left|\frac{21J_2R_P^2}{4a^2(1-e^2)^2}+\frac{6nL}{\Omega_P(1-e^2)^{3/2}}\right|.
\end{equation}
Inserting the numerical values for the various constants, the fractional correction to the winding rate is basically the same as the fractional correction to the precession rate:
\begin{equation}
\left|\frac{\partial \dot{\varpi}}{\partial r}\right|\sim3\times10^{-5}{\rm km^{-1}/day}\left|1-0.2\frac{(\Phi/-2{\rm V})}{(s/1 \mu{\rm m})^2}\right|.
\end{equation}
The pattern's estimated winding rate is 3.8\%$\pm$1.4\% slower than the expected value based on the planet's gravity field, which suggests that $(\Phi/-2 $V$)(s/1 \mu$m$)^{-2} = 0.19\pm0.07$, and that the effective average particle size is $s_{\rm eff} \simeq 2.3\pm0.5 {\rm \mu m}\sqrt{\Phi/(-2 {\rm V})}$,  comparable to the estimate of $s_{\rm diff}$ derived above. Of course, to evaluate whether these numbers are really consistent with each other will require a careful consideration of the various moments of ``realistic'' particle size distributions (i.e., distributions consistent with the rings' spectral and photometric properties), which is beyond the scope of this particular report. Still, these rough calculations indicate that size-dependent precession induced by Saturn's magnetic field is one potential explanation for the observed trends in the pattern's amplitude.

\subsection{What created the pattern?}

Extrapolating from the available observations,  the pattern should have had zero wavenumber on Julian Day 2455903$\pm$8, or December 7$\pm$8 of 2011. At that time, all the ring particles in this region would have aligned orbital pericenters. Thus it is reasonable to conclude that the event responsible for producing organized epicyclic motions in this region occurred around this same time. Since Cassini was in orbit around Saturn when the ring was disturbed, there is a chance that some instrument onboard the spacecraft might have observed phenomena related to this event. 

Before considering specific mechanisms for disturbing the ring, it is important to note that the magnitude of the event required to produce the pattern can be estimated by integrating the relevant perturbation equation for orbital eccentricities \citep{Burns76, Burns76err}. 
Assuming that a particle with mass $m$, radius $s$ and mass density $\rho_p$ was initially on a circular orbit, and that the perturbing force has a fixed magnitude $F_p$ and acted in a fixed inertial direction for a time $\Delta t$, the eccentricity induced by this force will be 
\begin{equation}
e= C n \frac{F_p \Delta t}{F_G},
\end{equation}
where $n=\sqrt{GM/a^3}$, $F_G=GMm/a^2=mn^2a$, and $C$ is a numerical constant of order unity. The net impulse required to produce a given eccentricity $e$ is therefore:
\begin{equation}
F_p \Delta t = \frac{e F_G}{Cn}=\frac{1}{C}{ae mn}=\frac{4\pi}{3C}ae n\rho_p s^3.
\end{equation}
The rapid attenuation of the pattern makes the initial eccentricities of the particles somewhat uncertain, but extrapolating from the observed trends, the original radial displacements are unlikely to exceed 100 km, so  we may take $ae \simeq 100$ km. Producing such an eccentricity would require each micron-sized water-ice rich ring particle orbiting 70,000 km  from Saturn's center to acquire an impulse of order $10^{-13}$ kg m/s.

Another way to quantify this disturbance is the total impulse that needs to be delivered into the entire ring. If the affected area of the ring has an optical depth $\tau$ and the disturbance extends over a region of width $w$ centered on semi-major axis $a$, then this total impulse is:
\begin{equation}
F_{tot} \Delta t =\frac{8\pi}{3C}ae  n\tau\rho_p s a w,
\end{equation}
The pattern of interest here extends over roughly 2000 km and appears across a region with an average optical depth of around $10^{-4}$. Assuming that the typical particles in this region are micron-sized and ice rich, the total impulse required to produce the observed pattern is of order $5\times10^9$ kg m/s. Note that since the eccentric spiral could extend outside the region between 68,000 and 70,000 km, where it would be difficult to detect against other ring structures, this should
be considered a lower limit on the required impulse.

The required momentum could have been imparted to the rings via collisions with a cloud of cometary debris. Indeed, vertical corrugations in Jupiter's rings were likely created by material from Shoemaker-Levy 9 crashing through the rings \citep{Showalter11}, and a similar collision is a favored explanation for the corrugations and eccentric spirals found elsewhere in Saturn's rings \citep{Hedman11, Hedman15}. Assuming the debris passed through the rings at close to the escape speed ($\sim 40$ km/s), the required impulse of at least $5\times10^9$ kg m/s could have been delivered into the rings by only 100,000 kg of material, equivalent to an ice-rich body a few meters across. However, to produce the observed spiral pattern this material would need to be a diffuse cloud of material rather than a single object, and so only a fraction of the mass passing through the rings will strike ring material. For a low optical depth ring and a diffuse cloud, the ratio of the amount of debris that strikes the ring is $\tau$ times the total mass passing through the rings, and so the total mass passing through the ring would need to be at least of order $10^9$ kg, equivalent to a solid object 60 meters wide. 

The small amount of material required to disturb the rings would be difficult to detect directly, but the debris that struck the rings could be part of a broader stream of material following the orbit of some more massive object, analogous to the meteor streams in the inner Solar System  \citep{Hedman11, Hedman15}. In this scenario, Cassini's cameras could have captured images of these impacts into the rings or the planet's atmosphere, but a preliminary inspection of the few available images of the rings and Saturn from this time failed to reveal any obvious evidence of impact phenomena. This is not strong evidence against this scenario because impacts into the rings are only likely to be visible for a limited range of viewing geometries that were not available during the relevant interval \citep{Chambers08, Tiscareno13}, and impact flashes in the atmosphere might also be too subtle to detect.  If the stream of cometary debris was sufficiently intense and extensive, then some of the material could have also struck Cassini. These impacts could potentially be detected by the  Cosmic Dust Analyzer (CDA) or the Radio and Plasma  Wave Spectrometer (RPWS) instruments onboard the spacecraft. Neither instrument has yet  reported an enhanced dust flux during the relevant epoch, but this lack of clear signals could potentially be explained by the extremely small size of the particles involved. Collisions with interplanetary debris can only efficiently disturb the ring if the debris particles are much smaller than the ring particles \citep{Hedman11}, and the particles visible in the inner D ring appear to be only a few microns across \citep{Hedman07}, so the debris responsible for producing the spiral pattern would need to be very small, probably sub-micron, grains. Such small particles would be more difficult for either CDA or RPWS to detect. 

Besides searching for evidence of a debris stream in the Cassini data, one can also explore whether Saturn was in a region of space where it might be likely to encounter cometary material. 
\citet{Hedman15} examined whether the ring-disturbing event in 1983 could have been caused by debris following the orbits of known objects, and found two Centaurs (Thereus and 2010 LJ109) with orbits that not only passed close to Saturn's location in 1983, but also were suitably oriented to deliver debris into the rings at the required angles to produce the observed epicyclic motions. A preliminary search for objects with orbits that came within 0.5 AU of Saturn in 2011 was conducted using orbital elements from the Minor Planet Catalog ({\tt www.minorplanetcenter.net/iau/MPCORB/MPCORB.dat}). Only one object was found to have a suitable orbit: 2007 DU112. Material approaching Saturn along this object's orbit would strike the rings at the fairly shallow angle of 30$^\circ$ (neglecting gravitational focusing effects), which would be marginally consistent with the pattern being a clear eccentric spiral without an obvious accompanying corrugation (see Appendix A). However, Saturn remained over 0.3 AU from 2007 DU112's orbit in 2011, twice its closest approach distance to Thereus' orbit in 1983. Thus there is not such a strong case that debris following the orbit of a known object could have generated the new pattern.


One potential alternative to a cometary impact is that the ring particles were perturbed by an electromagnetic phenomenon. Such a scenario is reasonable because the small particles that dominate the visible appearance of this ring can have large charge-to-mass ratios that make them particularly sensitive to non-gravitational forces. If we assume that the micron-sized grains in this ring region are charged to a few volts, then the typical charge on the grains will be of order $2\times10^{-16}$ C. The required impulse could then be delivered to the particles by an electric field
with a magnitude of  order 20 mV/m over a timescale of one orbital period. Smaller electric fields could deliver the required impulse if the duration of the disturbance is longer, but the perturbation will most efficiently alter the particle's orbits if it is applied over a time scale that is short compared to the local precession period, which is roughly two weeks for this part of D ring. Thus the electric field would probably need to be at least 0.4 mV/m to produce the observed pattern. Displacements in the microsignatures associated with Saturn's various moons indicate that there is an electric field in Saturn's magnetosphere pointing away from the Sun with a magnitude of order a few tenths of a mV/m \citep{Andri12, Andri14}. This field also appears to be time-variable \citep{Andri14}, so something could have happened in later 2011 that caused a sudden surge in this field's magnitude, imparting  the needed impulse into the rings.  More work is needed to ascertain whether the electric field surges  observed by Cassini well outside the main rings could also occur in the vicinity of the D ring, and whether sufficiently strong surges could have occurred in late 2011 or 1979.  Furthermore, the smaller-amplitude variations in the electric field could potentially influence the ring-particles' orbital evolution and thus affect the pattern's amplitude and wavelength, but such calculations are beyond the scope of this paper and thus must be left to a future work.

Alternatively, the required impulse could have been generated if the particles experienced a time-variable magnetic field as they orbited around the planet.  To produce the required impulse via the Lorentz force, the magnetic field would  need to vary  by a few hundred nT. Cassini magnetometer measurements  reveal not only that Saturn's magnetic field has azimuthal asymmetries, but also that the phase and amplitude of those asymmetries may change abruptly \citep{Provan13, Provan14}. However, none of the observed shifts in the magnetospheric oscillations appear to have occurred in late 2011. Furthermore, since these oscillations may involve current systems exterior to the planet itself \citep{Andrews12}, it is unclear if these asymmetries would be large enough to significantly perturb the D ring. 

\begin{figure}[tb]
\centerline{\resizebox{5in}{!}{\includegraphics{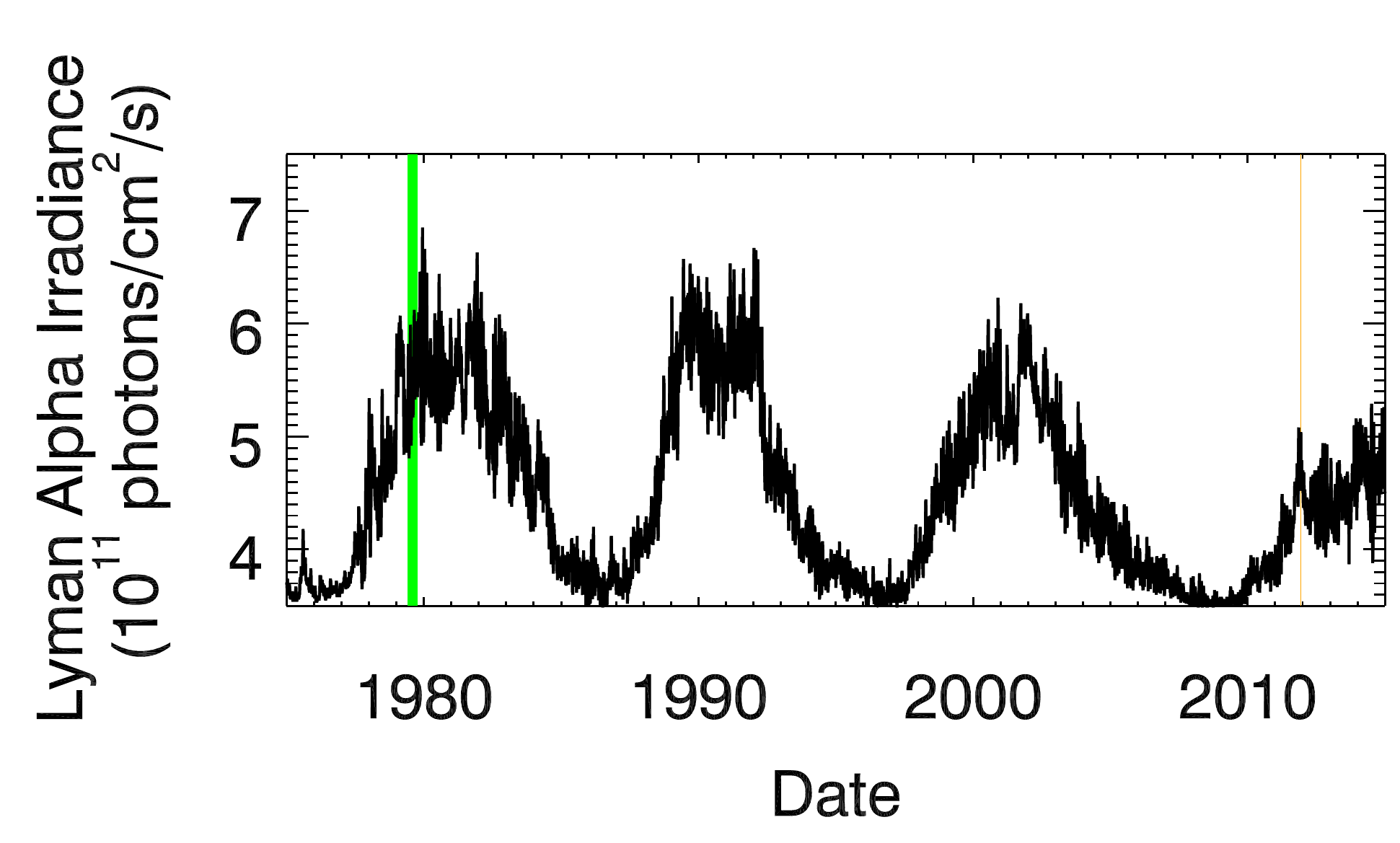}}}
\caption{Possible correlations between the ring-disturbing events and solar activity. The curve shows the Solar Lyman alpha flux composite time series from the LASP Interactive Solar Irradiance Datacenter ({\tt  http://lasp.colorado.edu/lisird}), with the approximate times of the ring-disturbing events in 2011 and 1979 indicated by colored bands. Note that both events occur close to narrow peaks in the flux near the beginning of solar maxima.}
\label{suntime}
\end{figure}

Finally, there are some intriguing similarities in Saturn's space environment during the two disturbance epochs of late 2011 and 1979. These two events are 32 years apart, which is not only close to one full Saturn year, but also three solar cycles. Thus both events occurred near Saturn's vernal equinox and around solar maximum. Indeed, as shown in Figure~\ref{suntime}, both events occurred close to narrow peaks in the solar ultraviolet flux where solar activity reached its highest levels during the appropriate maximum. This could imply that these events were somehow triggered by extreme levels of solar activity. There is evidence that the spikes in solar activity do correlate with abrupt changes in Saturn's magnetospheric structure, such as the radial extent of its electron radiation belts \citep{Roussos14}, and so these pulses of solar ultraviolet flux could potentially have generated the surges in the electric and magnetic fields discussed above. Alternatively, changes in the ultraviolet flux could change the temperature and density of Saturn's upper atmosphere \citep{Koskinen14}, changing the ionospheric drag felt by the D ring particles on the planet's dayside. In this context, it is interesting to note that the flux spikes were higher in 1979 than they were in 2011, which might be consistent with the larger amplitude of the patterns observed by Voyager.

\section{Summary}

The above analysis of the available imaging data provides the following insights into the structure that appeared in the D ring after 2011:
\begin{itemize}
\item The observed pattern was likely generated in late 2011 when some event gave the ring particles finite orbital eccentricities and aligned pericenters.
\item The organized epicyclic motions generated by this event have been steadily decaying away since they were first created.
\item  Particles of different sizes exhibit epicyclic motions with different amplitudes.
\end{itemize}
These data do not yet provide conclusive evidence for what created this pattern or what process is causing the pattern to dissipate so rapidly. Further investigations are therefore needed to provide answers to these questions.

\section*{Acknowledgements}
We thank the Imaging Team, the Cassini Project and NASA for providing the data used for this analysis. This work was funded by the Cassini Data Analysis Program Grant NNX12AC29G. MRS was also supported by the NASA Outer Planets Program Grant NNX14AO40G. We also want to thank E. Roussos, M. El Moutamid, J. Schmidt, P.D. Nicholson and J.A. Burns for useful conversations. In addition, we want to thank the reviewers for their very helpful suggestions regarding an earlier version of this manuscript.

\section*{Appendix A: Limits on a vertical corrugation.}

While the overall appearance of this new pattern is consistent with an eccentric spiral, a weak vertical corrugation could also be present in this region.  In principle the brightness variations due to a corrugation can be differentiated from those due to real optical depth patterns like an eccentric spiral because the intensity of the corrugation-induced patterns vary systematically with the observation geometry. Indeed, \citet{Hedman15} developed algorithms to isolate the signals due to vertical corrugations from those generated by eccentric spirals. These same tools were applied to the WAC images listed in Table~\ref{obstab} (the NAC images generally did not have sufficient azimuthal coverage or signal-to-noise for these algorithms to work). More specifically, for each image the region of the rings where the  parameter $|\cos\phi\cot B| < 0.5$ was considered ($B$ being the ring opening angle and $\phi$ being the azimuthal angle between the line of sight and the local radial direction). This analysis revealed a  longitude-dependent signal that, if it were due to a corrugation, would imply a corrugation amplitude of roughly one-half the amplitude of the eccentric spiral. 

However, this signal cannot be regarded as a positive detection of a vertical corrugation. Unlike the images considered in \citet{Hedman15}, the WAC images of the new pattern are all obtained at extremely high phase angles. In this viewing geometry, the rings' brightness is a very steep function of phase angle \citep{Hedman07}, and so there are detectable changes in the D-ring's apparent brightness across each image. These changes in brightness with phase angle could produce changes in the intensity of the pattern with longitude that the algorithms would mistake for  vertical structures. Addressing this issue would require detailed modeling of the ring-material's phase function, which is beyond the scope of this report. Thus this analysis merely indicates that the amplitude of any vertical corrugation in this region must be less than one-half the amplitude of the radial epicyclic motions producing the eccentric spiral.


\end{document}